%% file: main.tex
\documentclass[runningheads]{llncs}
\usepackage{color,soul}
\usepackage{tcolorbox}
\usepackage{xcolor}
\usepackage{graphicx}
\usepackage{algorithm}
\usepackage{algorithmic}
\usepackage{multicol}
\usepackage{multirow}
\usepackage{booktabs}
\usepackage{tabularx}
\usepackage{longtable}
\usepackage{array}
\usepackage{threeparttable}
\usepackage{colortbl}
\usepackage{enumitem}
\usepackage{hyperref} 
\usepackage{colortbl}
\usepackage{subcaption}
\usepackage{pifont}
\usepackage{amssymb}
\usepackage[pro]{fontawesome5}

\usepackage{marginnote}
\usepackage{xspace}

\usepackage{listings}[mathescape=true]
\definecolor{codegreen}{rgb}{0,0.6,0}
\definecolor{codegray}{rgb}{0.5,0.5,0.5}
\definecolor{codepurple}{rgb}{0.58,0,0.82}
\definecolor{backcolour}{rgb}{0.95,0.95,0.92}
\lstdefinestyle{mystyle}{
    backgroundcolor=\color{backcolour},   
    commentstyle=\color{codegreen},
    keywordstyle=\color{magenta},
    numberstyle=\tiny\color{codegray},
    stringstyle=\color{codepurple},
    basicstyle=\ttfamily\footnotesize,
    breakatwhitespace=false,         
    breaklines=true,                 
    captionpos=b,                    
    keepspaces=true,                 
    numbers=left,                    
    numbersep=5pt,                  
    showspaces=false,                
    showstringspaces=false,
    showtabs=false,                  
    tabsize=2
}
\newcommand{\one}{\textcolor{black}{\large\textbf{\ding{172}}}\xspace}
\newcommand{\two}{\textcolor{black}{\large\textbf{\ding{173}}}\xspace}
\newcommand{\three}{\textcolor{black}{\large\textbf{\ding{174}}}\xspace}

\newcommand{\rev}[1]{\textcolor{black}{#1}}

\tcbuselibrary{theorems}
\newtcbtheorem{good-to-know}{Good-to-know \faIcon{bullhorn}}
{colback=yellow!5,colframe=blue!35!black,fonttitle=\bfseries}{th}
\newtcbtheorem{tip}{Tip \faIcon{lightbulb}}
{colback=yellow!5,colframe=blue!35!black,fonttitle=\bfseries}{th}
\newtcbtheorem{Definition}{Definition  \faIcon{info-circle}}
{colback=magenta!5,colframe=violet,fonttitle=\bfseries}{th}
\definecolor{lava}{rgb}{0.81, 0.06, 0.13}\newtcbtheorem{challenge}{Challenge  \faIcon{mountain}}
{colback=black!5,colframe=lava,fonttitle=\bfseries}{th}

\newcommand{\sectopic}[1]{\vspace{0em}\par\noindent{\textit{\bfseries #1}}}

\def\BibTeX{{\rm B\kern-.05em{\sc i\kern-.025em b}\kern-.08em
T\kern-.1667em\lower.7ex\hbox{E}\kern-.125emX}}
\begin{document}
\title{Chapter 11: Legal Requirements Analysis: A Regulatory Compliance Perspective}
%
%
\author{Sallam Abualhaija\inst{1}\orcidID{0000-0001-6095-447X} \and
Marcello Ceci\inst{1}\orcidID{0000-0003-3800-0906} \and 
Lionel Briand\inst{2,3}\orcidID{0000-0002-1393-1010}\thanks{Part of this work was done while the author was affiliated with the University of Luxembourg, Luxembourg} }
\authorrunning{S. Abualhaija et al.}
%
\institute{SnT, University of Luxembourg, Luxembourg
\email{\{sallam.abualhaija,marcello.ceci\}@uni.lu}\\ \and
School of EECS, University of Ottawa, Canada\\
\email{lbriand@uottawa.ca}
\and
Lero SFI centre for Software Research and University of Limerick, Ireland\\
\email{lionel.briand@lero.ie}
}
\maketitle              
\begin{abstract}
\input{abstract}

\end{abstract}
\newpage
\input{introduction}

\input{elicitation}

\input{compliance}
\input{challenges}

\input{conclusion}

\bibliographystyle{splncs04}
\bibliography{references}
%




\end{document}

%% file: abstract.tex
Modern software has been an integral part of everyday activities in many disciplines and application contexts. 
Introducing intelligent automation by leveraging artificial intelligence (AI) led to breakthroughs in many fields. The effectiveness of AI can be attributed to several factors, among which is the increasing availability of data. 
Regulations such as the general data protection regulation (GDPR) in the European Union (EU) are introduced to ensure the protection of personal data. 
Software systems that collect, process, or share personal data are subject to compliance with such regulations. Developing compliant software depends heavily on addressing legal requirements stipulated in applicable regulations, a central activity in the requirements engineering (RE) phase  
of the software development process. 
RE is concerned with specifying and maintaining requirements of a system-to-be, including legal requirements. Legal agreements which describe the policies organizations implement for processing personal data can provide an additional source to regulations for eliciting legal requirements.  
In this chapter, we explore a variety of methods for analyzing legal requirements and exemplify them on GDPR. Specifically, we describe possible alternatives for creating machine-analyzable representations from regulations, survey the existing automated means for enabling compliance verification against regulations, and further reflect on the current challenges of legal requirements analysis.

%% file: introduction.tex
\section{Introduction}~\label{sec:introduction}

Modern software systems are becoming increasingly large, complex, and most importantly data-driven. Recent notable breakthroughs are resulting in technologies based on artificial intelligence (AI)  increasingly mastering different dimensions of human communication, including text understanding and generation, image and sound processing. AI-enabled software is being integrated in various application domains, e.g., healthcare~\cite{Caruana:15}, transport~\cite{Cui:19}, manufacturing~\cite{Wang:18}, and finance~\cite{Huynh:20}. In tandem with the adoption such systems into our societies, regulations are being continuously introduced to ensure that the software development lifecycle is lawful, ethical, and robust~\cite{AIEthics}. 
Developing software systems that are compliant with the applicable law and regulations is one of the core interests in the requirements engineering (RE) field. Legal requirements analysis entails creating  machine-analyzable representations of legal text based on which automated analysis technologies can be developed. In this chapter, we focus on building automated approaches for enabling compliance checking, a major concern for requirements engineers.  

The general data protection regulation (GDPR) is considered the benchmark for data protection and privacy standards~\cite{GDPR}. Since its inception in 2018, it has been extensively studied for various purposes, including the elicitation of legal requirements related to privacy and data processing-related concerns which must be addressed in software systems.   
Given the far-reaching impact that GDPR has beyond the EU, we use it to  exemplify the various methods described in this chapter. Any organization operating in the EU, irrespective of where it is located, 
must comply with the GDPR provisions, or face costly fines
\footnote{\href{https://edpb.europa.eu/news/news/2023/12-billion-euro-fine-facebook-result-edpb-binding-decision_en}{1.2 billion euro fine for Facebook as a result of breaching GDPR}}. Analyzing the legal requirements stipulated in GDPR is thus an essential prerequisite for developing compliant software systems. We note that the methods we explain in this chapter are not specific to GDPR. Using GDPR as an example helps first illustrate how the methods work. 
The same methods can then be adapted and applied to other regulations.   

\sectopic{Running Example. } Consider the following example (adapted from existing RE work~\cite{abualhaija2022coreqqa}). 
In relation with GDPR, a certain procedure needs to be followed for handling personal data breaches. Concretely, Article 33 in the GDPR contains, among other things, the following provision: 

\begin{tcolorbox}[arc=0mm,width=\columnwidth,
                  top=1mm,left=1mm,  right=1mm, bottom=1mm,
                  boxrule=1pt] 
\begin{example} \label{eg1}
``\emph{In the case of a personal data breach, the controller shall without undue delay and, where feasible, not later than 72 hours after having become aware of it, notify the personal data breach to the supervisory authority competent in accordance with Article 55, unless the personal data breach is unlikely to result in a risk to the rights and freedoms of natural persons. Where the notification to the supervisory authority is not made within 72 hours, it shall be accompanied by reasons for the delay.}''
\end{example}
\end{tcolorbox}

Analyzing, understanding, and representing this provision will be discussed in the following sections. Generally speaking, from an RE standpoint, this provision will be translated into a set of actions that describe how the system deals with personal data breaches. Requirements engineers will specify corresponding requirements that aim at satisfying this provision. Examples of such requirements are listed below: 

\begin{itemize}
 \item[] \textbf{REQ1.} If a data breach is identified on the \texttt{SYSTEM} server, the \texttt{NotifyService} shall be activated.
 \item[] \textbf{REQ2.} The \texttt{NotifyService} shall send an email to the \textit{Chief Information Officer} notifying the breach, within 72 hours of its occurrence. 
 \item[] \textbf{REQ3.} Upon opening the email on the receiver's side, a automated notification about the read receipt shall be  collected and stored on the server.
\end{itemize}

\sectopic{Terminology. } We will now define the terminology that will be used throughout the chapter:
\begin{itemize} [wide]
    \item \textit{Legal requirement} is defined as a requirement 
    that has its source in a legislative document. Requirements are derived from the norms expressed by legal provisions (see Section \ref{sec:representation} for a definition of \emph{provision} and \emph{norm}).
    \item \textit{Compliance checking} means establishing the compliance of a software, i.e., the relationship between the specification of the system-to-be and the set of relevant legal norms~\cite{jureta21}. It thus concerns the relationship between the requirement documentation and the norms expressed by the law.
    \item \textit{Breach}, from an RE perspective, is the situation when a legal requirement is not fulfilled by the system. We will introduce in Section~\ref{sec:representation} the concept of breach from a legal theoretical perspective.
\end{itemize}

\sectopic{Scenarios. } There are essentially two RE scenarios in which we can analyze legal requirements: we can either perform question  answering on the legal text directly (also called ``advanced legal information retrieval'' \cite{Francesconi19}) or we can build a machine analyzable representation of the legal text for the purpose of developing automated support for legal compliance checking. 

\sectopic{Structure. } 
\begin{figure}[!t]
\centering
\includegraphics[width=0.9\linewidth] {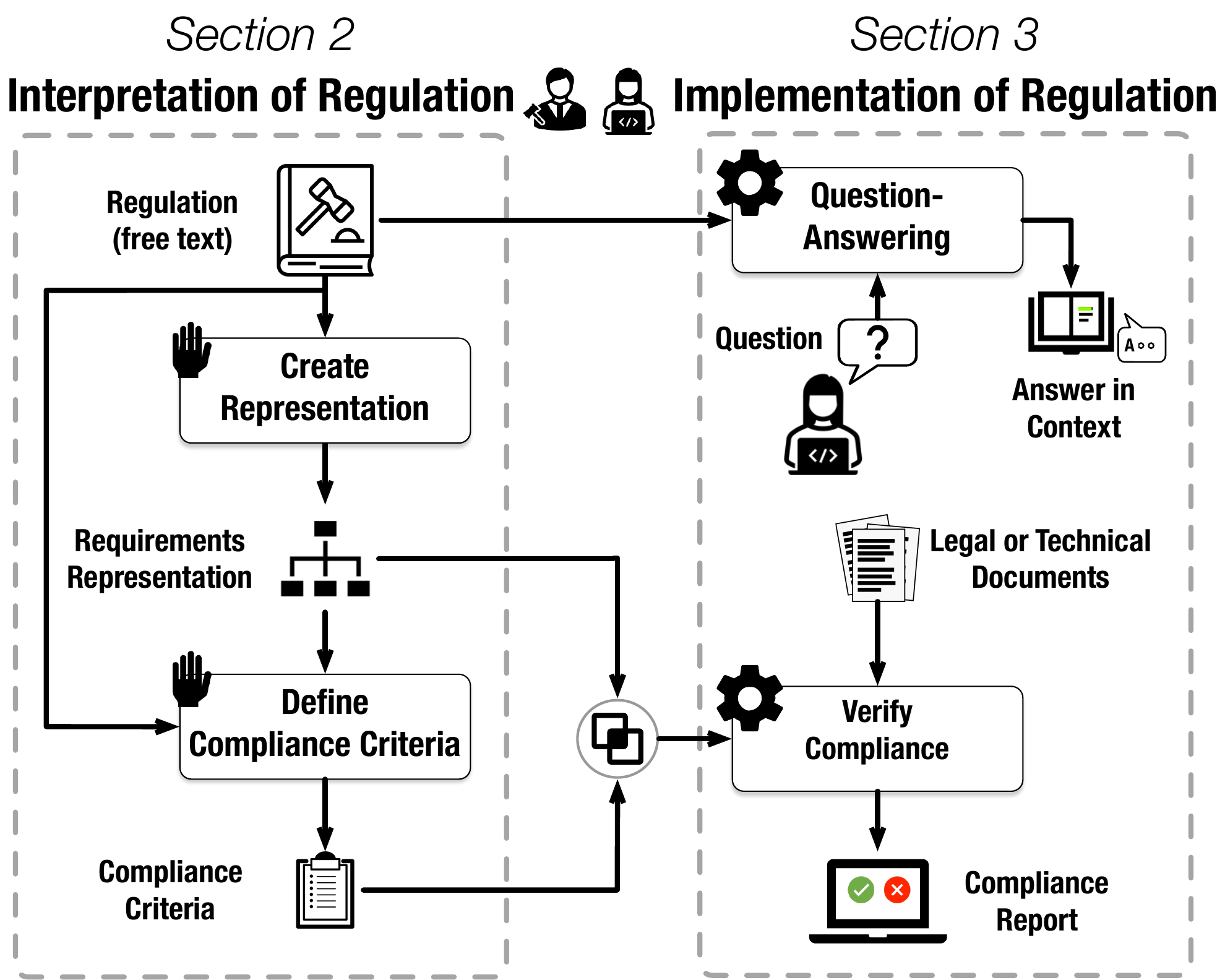}
\caption{Chapter Overview.} \label{fig:overview}
\vspace{-2em}
\end{figure}
Fig.~\ref{fig:overview} shows an overview of the chapter content. 
On the left-hand side, we explain how to interpret the regulation 
to enable compliance checking. To do so, requirements engineers need to create machine-analyzable representations and define compliance criteria, two activities that are often done manually in collaboration with legal experts.  
On the right-hand side, we explain how to analyze legal requirements by retrieving compliance-relevant information from the free text of the regulation through question-answering as well as  building automated approaches for checking compliance according to the machine-analyzable representation of the regulation and the defined compliance criteria.

%% file: elicitation.tex
\section{From Regulations to Representation}
\label{sec:representation}

Legal requirements pertain to the domain of law and are derived from legal orders, i.e., speech acts with descriptive or prescriptive functions emanated by an authority~\cite{Francesconi19}. 
In every such legal order we can identify two aspects: a sequence of words structured in sentences, called \textit{Provision}, and the meaning expressed by those words, called \textit{Norm}, which results from the interpretative process~\cite{kelsen91}. Norms are therefore the rules expressed by provisions. These norms (more specifically, regulative norms -- see below) describe a desired reality to which the facts are required to conform. The matching of any facts to these norms is called \textit{compliance}, while any discrepancy among the two is called a \textit{breach}\footnote{A breach is defined here as ``a failure to do what is required by a law, an agreement, or a duty, i.e., failure to act in a required or promised way'' \cite{brit}.}.

To obtain a machine-analyzable representation of legal requirements, we have the option to work towards a general and universally applicable representation of provisions and norms (which is an objective of the discipline of \emph{legal informatics}) or rather create \emph{ad hoc} solutions for representing regulations in light of a specific use case, which describes a business activity within a legal domain or disciplined by a specific regulation. 
While formal models aimed at establishing a standard general representation of the law are investigated in the literature~\cite{BENCHCAPON2003,jureta21}, they are not yet advanced enough for widespread application, and are hard to apply in practice. 
Following common practices in the RE literature~\cite{kiya08,Aramal21,Torre21}, this chapter 
focuses on representing norms that are applicable at one specific point in time in a specific jurisdiction. 
Such a representation reduces the complexity of legal knowledge to be represented, especially with regard to 
temporal parameters and 
regulatory change management. 

\begin{good-to-know*}{}
\textbf{Legal Informatics} is an interdisciplinary branch of legal science and informatics, striving to complement the traditional legal perspective with perspectives from the field of information sciences (system theory, computer science, communication theory, information security theory, cognitive science, and library science)~\cite{seipel2004}. Its areas of research include the representation of legal concepts, legal norms, and legal reasoning in a machine-readable and machine-computable way. 
\end{good-to-know*}

When eliciting legal requirements, 
the first step consists in identifying the different legal norms expressed in the regulation. Following are the main \textbf{types of norms}, as identified in the literature~\cite{Sleimi18}:
\begin{itemize}
    \item \emph{Obligations and prohibitions} (called \emph{regulative norms}) 
    are two complementary ways of representing the conditions for compliance. For obligations, 
    the facts of the case in point must match those described by the legal norm. 
    For prohibitions, the contrary is true, i.e., a breach occurs when the facts match the legal norm. Regulative norms can also be expressed in the legal text as rights or powers, following Hohfeld's classification~\cite{hohfeld913}. However, in order to correctly capture legal requirements, rights and powers must first be interpreted, i.e., translated into corresponding obligations or prohibitions~\cite{Sleimi:19}.
    \item \emph{Definitions} pertain to another type of legal norms, called \emph{constitutive norms}, which do not impose a behavior but rather define new legal entities~\cite{searle1969}. 
    \item Finally, \emph{penalties} attach a legal effect (called sanction) to the breach of a regulative norm. Because it triggers when a rule is breached, the effects and semantics of penalties are closely tied to those of regulative norms~\cite{jones02}. 
    
\end{itemize}

\begin{Definition*}

According to legal theory, \textbf{regulative norms} are the result of directive acts. Regulative norms regulate behaviors by introducing
deontic modalities (e.g., it is obligatory that…, it is prohibited that…, it is permitted that…) for their addressees. Their application is 
conditional, as norms specify their preconditions.

\textbf{Constitutive norms}, on the other hand, are the result of declarative acts. These norms introduce new abstract classifications of existing facts and entities, e.g., concepts such as marriage, money, private property. 
Unlike regulative rules,
constitutive norms have no deontic content, i.e., they do not introduce obligations, prohibitions or permissions. 
For more information on constitutive rules in the context of RE, we refer the reader to Sleimi et al.~\cite{searle2018constitutive}.
\end{Definition*}

For instance, in the legal provision of \textit{Example}~\ref{eg1} (Section~\ref{sec:introduction}) we can identify 
the following norms:

\begin{enumerate}[wide]
\item The obligation to notify within 72 hours any personal data breach that is likely to result in a risk for persons;
\item The obligation to accompany the notification with the reasons for the delay in case the notification is not done within 72 hours.
\end{enumerate}

Representing legal provisions as norms (similar to what we did above) is a complex activity, since 
it involves 
\textbf{legal interpretation} and unavoidably requires the disambiguation of legal concepts~\cite{lopez20}.
The representation of concepts and even norms themselves thus varies according to 
the use case (e.g. law application in a tribunal, legal counseling, requirements elicitation) and the perspective taken (e.g, the regulator, the holder of a right or of a legitimate interest). 
In RE, the focus is on 
those legal norms that describe a desired reality which pertains to what can be automated,  
particularly those legal norms that regulate actions or activities performed by software, as these are the rules that have to be taken into account during the development of new --- or for checking the compliance of existing --- software. 
This implies that the rules would be expressed as addressing a software system (``\emph{the software system} shall...'').
At the same time, the compliance space is restricted to the aspects of the software system that are the object of normative intervention. For example, the ``facts'' to which legal requirements are applied are artifacts rather than activities or events.

The complexity of legal knowledge and the main challenges to its representation will be elaborated in Section \ref{challenges}.


\subsection{Machine-analyzable Representation of Regulation} \label{regrep}

Representing regulations involves extracting the \textit{structural metadata} regarding the document and the \textit{domain concepts} (i.e., elements of the desired reality) described therein. The \textit{norms} expressed by the regulations and the \textit{legal concepts} are extracted as compliance criteria (as we explain in Section~\ref{criteria}). The representations we discuss in this section serve as enablers for the NLP technologies that we explain in Section~\ref{sec:compliance}. Such representations are typically created manually for a particular application context with the intensive involvement of legal experts. 

\textbf{Structural metadata} describing the legal document are:

\begin{itemize}
    \item \emph{Document ID} -- Metadata identifying the document. 
    In \textit{Example}~1, the official name of the document containing this provision 
    can be 
    represented by the CELEX\footnote{A CELEX number is a unique identifier assigned to a document, used in EUR-Lex. See: https://eur-lex.europa.eu/content/help/eurlex-content/celex-number. } number: ``32016R0679'' or the European Legislation Identifier (ELI): \href{http://data.europa.eu/eli/reg/2016/679/oj}{http://data.europa.eu/eli/reg/2016/679/oj}.
    \item \emph{Hierarchical structure} -- Metadata identifying the structure of the document. In our running example, the legal statement corresponds to Article 33 in the hierarchical structure of document. 
    A standard representation of the structure of the regulation is essential for unequivocally referring to any rule. These metadata are also fundamental for regulatory change management (i.e., determine how applicable norms change over time). 
    \item \emph{Lifecycle, publication, validity} -- Metadata pertaining to the lifecycle of the document (e.g. the steps of discussion of a law in parliament, the publication date and location, and the date of entry into force) are fundamental to determine the applicability of legal norms through time.
    \item \emph{Cross references} -- Metadata describing references contained in the law to other parts of the same law or other laws or legal instruments (e.g., ``competent in accordance with Article 55'' in \textit{Example~1}). These metadata are important for 
    determining the exact content of the norms and 
    the interrelation between them.
    \item \emph{Modifications} -- Metadata describing the effect of those provisions that modify the text of other laws (e.g., ``in Article 12, the third paragraph is replaced by the following: [...]'') are fundamental for regulatory change management. 
\end{itemize}

Structural metadata are often available from the issuing institutions, using XML-based standards in legal portals, e.g. in UK\footnote{https://www.legislation.gov.uk/developer/formats/xml}; 
in the EU, the EUR-Lex website contains legislation together with a wide array of structural metadata~\cite{francesconi18}. 
Since they come from official issuing authorities, these are authoritative metadata and constitute precious sources for the automation of legal processes.  



The \textbf{domain concepts} 
contained in legal norms can be extracted to build reference models of the legal domain.  Several existing approaches have been proposed in the literature~\cite{Torre20,ceci2016,Anton09,siena2012capturing} to extract entities and relations from legal texts in order to build a model or ontology. 
Domain concepts are distinguished from \emph{legal concepts} (such as ``shall'' and ``unless'' in the running example) which are related to legal subjects, preconditions, and the concepts of breach and validity. Legal concepts should not be \emph{extracted} from the text but rather \emph{mapped} from a reference model of the law to enable the extraction of compliance criteria. 

Domain concepts are represented in domain models. The basic elements of a domain model include entities, relations, and attributes, defined as follows:

\begin{itemize}
    \item \emph{Entities} represent single creatures, artifacts or events. They can be hierarchically structured, e.g., \textit{Student} is a subconcept of \textit{Person}.
    \item \emph{Relations} represent interactions between two entities, e.g., Student \emph{enrolls in} Course.
    \item \emph{Attributes} represent characteristics of entities. They are attached to single entities and accompanied by a value of varying formats, e.g., an integer (Student \emph{has\_id=``005849''}) or a boolean (Student \textit{is\_enrolled=``true''}).
\end{itemize}

Domain concepts can be represented in different formats, elaborated next. 

\sectopic{Conceptual models. } A generic \textbf{conceptual model} defines the relevant domain concepts in a regulation and is usually built in collaboration with domain experts~\cite{kiya08,amaral20,Torre21}. This collaboration is necessary, as mentioned earlier, since the representation implies the interpretation of the provisions.  The approach may be more or less adherent to the text of the law, depending on the purpose of the model. For building the conceptual model, three types of coding can be subsequently applied~\cite{saldana09}: \emph{in-vivo coding} identifies the main concepts in the legal text to create codes, e.g., ``controller'' and ``personal data breach'' in \textit{Example}~\ref{eg1}; 
\emph{hypothesis coding} applies a predefined set of codes (obtained with in-vivo coding) to actual data to assess whether the codes and their level of abstraction are sufficient for the analysis (e.g., identifying ``controller'' and ``personal data breach'' in some privacy policies); 
\emph{subcoding} uses sub-codes as a second-order tag assigned after a primary code to enrich the specificity of the obtained metadata, e.g., ``within 72 hours'' can be added as specialization of ``notification time''. 

\begin{figure}[t!]
\centering
\begin{subfigure}{.38\textwidth}
  \centering
    \includegraphics[width=0.9\linewidth] {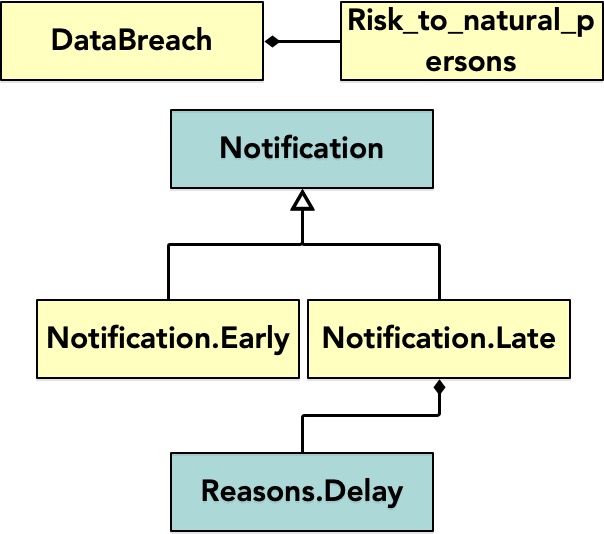}
\vspace*{-.2em}
    \caption{A simple conceptual model.}
    \label{fig:model2}
\end{subfigure}%
\begin{subfigure}{.62\textwidth}
  \centering
    \includegraphics[width=0.9\linewidth] {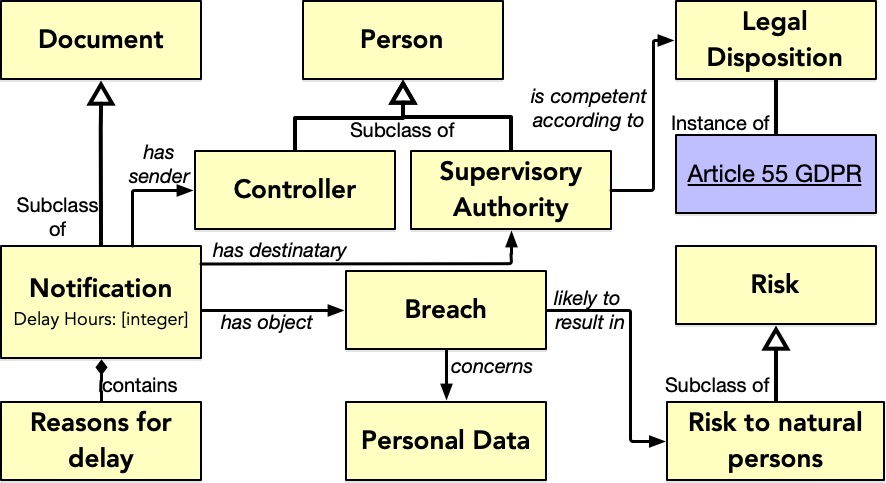}
    \vspace*{-.2em}

    \caption{A more complex conceptual model.} 
    \label{fig:ontology}
\end{subfigure}
\caption{Examples of conceptual models for the legal provision of \textit{Example}~\ref{eg1}.}
\label{fig:representation}
\vspace{-1em}
\end{figure}
A conceptual model is different from a \textbf{taxonomy}, i.e., a scheme of terminological and hierarchical classification in which terms are organized into groups or types. While taxonomies classify terms, conceptual models represent the concepts expressed by those terms. Unlike conceptual models which are specific to the application context, taxonomies are less application-specific. However, the reusability of taxonomies is still limited  by the lack of formally defined syntax and semantics. 
Taxonomies are used in the context of RE 
to describe standard terminology and as a resource to be used to build or expand conceptual models. For example, the hierarchy ``foreign military authority'' \textit{is-a} ``foreign government authority'' \textit{is-a} ``government authority'' can be a part of a taxonomy~\cite{breaux08}. 

\sectopic{Ontologies. }    
An \textbf{ontology} is a description (formal specification) of a conceptualization, i.e., of the concepts and relationships that can formally exist for an agent or a community of agents~\cite{gruber95}. 
Ontologies are similar in nature to conceptual models, but they differ in the fact that the definitions of the concepts include information about their meaning and constraints on their logically consistent application~\cite{gruber09}. 
Ontologies are  subject to a strict logic formalism, which on one hand makes them more difficult to build, but on the other hand provides the possibility to automatically validate the consistency of the expressed knowledge, and to use semantic reasoners to infer new knowledge. 
Ontologies are also typically specified in languages that allow abstraction away from data structures and implementation strategies. 
They are thus used for integrating heterogeneous databases, enabling interoperability among disparate systems. 
While building an ontology, it is possible to reuse existing higher-level ontologies in the context of the \textit{semantic web}, expressing the semantic relation between the different representations in a machine- and human-readable way. Ontologies have been used in the RE literature in various contexts~\cite{mosquera:23,dermeval2016}.

\begin{Definition*}{}

The \textbf{Semantic Web}~\cite{bonatti2006,henze2004} is an initiative undertaken by the W3C 
whose purpose is to increase the machine-readability of information available on the web by using various solutions including markup languages and formal representations and logic formalisations~\cite{Antoniou12}. Examples of W3C standards 
include RDF\footnote{The \textbf{Resource Description Framework (RDF)} is a World Wide Web Consortium (W3C) standard originally designed as a data model for metadata. It has come to be used as a general method for description and exchange of graph data~\cite{decker2000semantic}. RDF provides a variety of syntax notations and data serialization formats.
 }, OWL, and SHACL.
The Web Ontology Language (\textbf{OWL}) is a family of knowledge representation languages 
characterized by formal semantics. They are built upon the RDF standard 
and can be expressed in Extensible Markup Language (\textbf{XML}). 
\end{Definition*}

The standard W3C language used for representing ontologies is OWL 2. 
The most important foundational ontology for the law is LKIF (Legal Knowledge Interchange Format) \cite{hoekstra09}. PrOnto~\cite{palmirani18b} is an example of operational ontology in the domain of data privacy.


    
    


To illustrate the various representations discussed above, let us consider again \textit{Example}~\ref{eg1}. 
We can represent the example through a model expressed as a class diagram, as shown in Fig.~\ref{fig:representation}. The model can be rather simple (see Fig.~\ref{fig:model2}), or more complex (Fig.~\ref{fig:ontology}). The choice depends on factors such as the need for reusability of the model, the existence of pre-existing data, and the need for automation in the management of data. For example, if in our application context we do not handle notifications to authorities other than the competent supervisor, 
the representation of \emph{Supervisory Authority} in Fig.~\ref{fig:ontology} is not required and can be therefore removed. In this case, the representation of ~\ref{fig:model2} can be adopted 
where the model is simpler and easier to understand and use. The same provision can be represented in different formats. For example, the model in Fig.~\ref{fig:model2} can be expressed in text format as follows:

\begin{itemize}
    \item \textbf{Events}:DataBreach; Notification.Early; Notification.Late
    \item \textbf{Entities}: Reasons.Delay; Risk\_to\_natural\_person

\end{itemize}

We can also represent the same provision 
in an ontology. Here is 
an excerpt of an ontology representing the model in Fig.~\ref{fig:ontology} in OWL/XML syntax, reusing concepts from the LKIF-Core ontology\footnote{http://graph-data-model.infinitech-h2020.eu/content/lkif-docs/index-en.html}:

\renewcommand{\lstlistingname}{Representation \faIcon{code}}
\begin{lstlisting}[language=XML, label={lst:owlr}, caption=Entities Descriptions in OWL/XML, mathescape=true]
<! -- Description of the entities "Notification" and "SupervisoryAuthority" as subconcepts of "Document" and "Person" from the LKIF-Core ontology. -->
    <Declaration>
        <Class IRI="#Notification"/>
    </Declaration>
  <SubClassOf>
        <Class IRI="#Notification"/>
        <Class IRI="http://www.estrellaproject.org/lkif-core/expression.owl#Document"/>
    </SubClassOf>
        <Declaration>
        <Class IRI="#SupervisoryAuthority"/>
    </Declaration>
  <SubClassOf>
        <Class IRI="#SupervisoryAuthority"/>
        <Class IRI="http://www.estrellaproject.org/lkif-core/expression.owl#Person"/>
    </SubClassOf>
<! -- description of the object property "has_destinatary" linking a "Notification" and a "SupervisoryAuthority". -->
    <ObjectPropertyDomain>
        <ObjectProperty IRI="#has_destinatary"/>
        <Class IRI="#Notification"/>
    </ObjectPropertyDomain>
    <ObjectPropertyRange>
        <ObjectProperty IRI="#has_destinatary"/>
        <Class IRI="#SupervisoryAuthority"/>
    </ObjectPropertyRange>
 
\end{lstlisting}

\begin{tip*}{}
The choice among different representations depends on the purpose for which the model is intended to be used, which in turn impacts the level of interpretation of the regulation as well as the abstract formalization required. 
Taxonomies are useful to create hierarchical classifications of terms. Conceptual models are beneficial to represent concepts and relations within a specific application context. Compared to conceptual models, ontologies are often preferred when representing heterogeneous data (i.e., data from multiple sources). Ontologies are also more standardized, e.g.,
the concept of ``document'' as defined in an ontology should have 
a logical relation with a standard representation of ``document'', e.g., in a foundational ontology such as LKIF, whereas in a conceptual model ``document'' is only defined in the domain or use case for which the model is created. This way, a machine can understand what is defined in an ontology, properly capture the complexity of the law highlighted in this section, and further enable  new knowledge inference and interoperability. 

\end{tip*}
Together, the information provided by structural metadata and domain concepts can go to great extents towards supporting compliance checking. As we will learn in Section~\ref{sec:compliance}, automated support to  check data or text against legal requirements can make use of these elements to correctly evaluate the case in point in the light of the applicable norms.

\subsection{Definition of Compliance Criteria} \label{criteria}
\renewcommand{\lstlistingname}{Rule \faIcon{code}}

\textbf{Compliance criteria} are the rules that guide the compliance checking process. They express the modal restrictions introduced by the norm on the domain concepts (entities and relations) of a given conceptual model. 
%
To represent these rules we can use a natural language, tailored for the specific use case (e.g., compliance of a privacy policy) or a more generic, formal and thus universally-applicable logical formalism.

We discuss below different languages 
for specifying the compliance criteria derived from \textit{Example}~\ref{eg1}. They differ in terms of format, granularity of representation, and degree of formalization. 

\sectopic{(1) Natural-language rules:} Capturing compliance criteria as natural language statements is a widespread practice in RE field, often performed in close collaboration with domain experts~\cite{kiya08,amaral20,Torre21}.

\begin{tcolorbox}[arc=0mm,width=\columnwidth,
                  top=1mm,left=1mm,  right=1mm, bottom=1mm,
                  boxrule=1pt,colback=white,colframe=black]  
    \begin{enumerate}
    \item If a personal data breach is likely to result in a risk to natural persons, the controller shall notify the breach to the  supervisory authority within 72 hours from knowing the breach; 
    \item When the breach is notified after the time limit, the controller shall indicate the reasons of the delay.
\end{enumerate}
\end{tcolorbox}
\sectopic{(2) Template-based:}
%
The same statements can be structured according to a predefined template. Each criterion is composed of two parts, namely \textit{a precondition} (if any) which states the condition whose truth value needs to be checked to activate the criterion, and \textit{a postcondition} asserting the action required when the truth value of the precondition is \textit{TRUE}. One can use a template, e.g.,  IF [\textit{precondition}],
THEN $<$\textit{postcondition}$>$ to specify the  compliance criteria. 
Below is an example of template-based requirements based on the model of Fig.~\ref{fig:model2}:

\begin{tcolorbox}[arc=0mm,width=\columnwidth,
                  top=1mm,left=1mm,  right=1mm, bottom=1mm,
                  boxrule=1pt,colback=white,colframe=black]  
\begin{enumerate}
    \item IF [\textit{DataBreach.Risk\_to\_natural\_person}], THEN $<$\textit{Notification.Early}$>$; 
    \item IF [\textit{Notification.Late}] THEN $<$\textit{Reasons.Delay}$>$.
\end{enumerate}
\end{tcolorbox}

\begin{figure}[H]
\centering
\includegraphics[width=.8\linewidth] {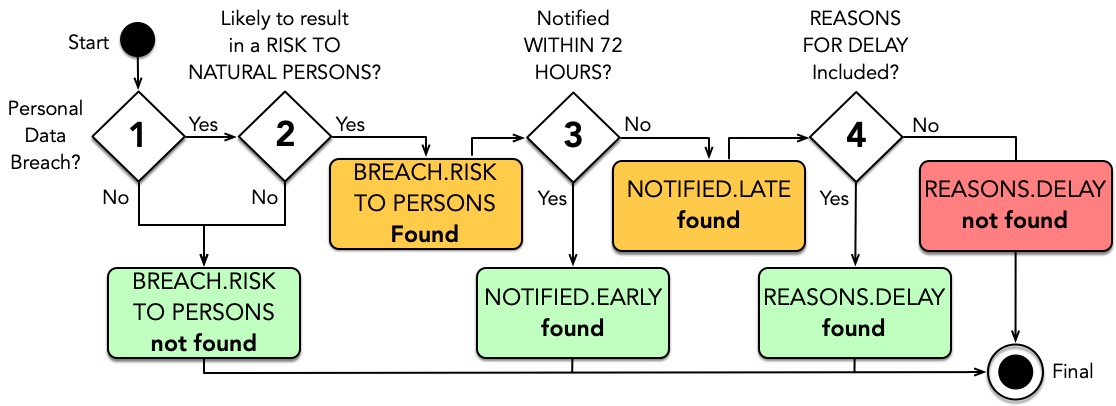}
\caption{Activity diagram for the compliance criteria of \textit{Example}~\ref{eg1}.} \label{fig:ad}
\vspace{-2em}
\end{figure}

\sectopic{(3) Activity Diagrams:} Activity diagrams such as the one in Fig.~\ref{fig:ad} are representations of workflow that, given basic training, can be understood also by legal experts \cite{soltana18}. 

\sectopic{(4) Formal logic:} 
%
\textit{Logical forms} represent 
a precisely-specified semantic version of a statement in a formal system. They are used for the representation of knowledge, for example in an ontology.
The following formula shows one logical form for the same provision using deontic logics, where the part in \textcolor{blue}{blue} corresponds to the text span ``unless the personal data breach is unlikely to result in a risk to the rights and freedoms of natural persons'', and the part in \textcolor{codepurple}{purple} corresponds to ``Where the notification to the supervisory authority is not made within 72 hours, it shall be accompanied by reasons for the delay.''

\begin{tcolorbox}[arc=0mm,width=\columnwidth,
                  top=1mm,left=1mm,  right=1mm, bottom=1mm,
                  boxrule=1pt,colback=white,colframe=black]  
$\exists_{b,d,a,c,r}~breach(B)~\land~data(D)~\land~authority (A)~\land~contoller (C)~\land risk(R)~\land \textcolor{blue}{concerns(B,D) \land \neg unlikely(B,R)} \land competent(A) \implies \\ $[O]$ \exists_n notification(N) \land performs(C,N) \land concerns(N,B) \\ \land destinatary(N,A) \land (delay(N,<72h) \textcolor{codepurple}{\lor (\exists_e~reasons(E) \land contains(N,E))})$
\end{tcolorbox}
\begin{Definition*}
    
    \textbf{Deontic logic} 
    is a formal system that attempts to capture the essential logical features of obligation, permission, and related concepts. Typically, deontic logic is formalized as a modal logic~\cite{vonwright51}, and it uses ${\displaystyle [O]A}$ to mean ``it is obligatory that A'' (or ``it ought to be (the case) that A''), and ${\displaystyle [P]A}$ to mean ``it is permitted (or permissible) that A'', which is defined as ${\displaystyle [P]A\equiv \neg [O]\neg A}$, where $\neg$ denotes a negation. 
    \textit{\textbf{Tip:} Can you determine why the negation is used twice in ${\displaystyle [P]A\equiv \neg [O]\neg A}$? }

    While in common logics we compare statements in terms of truth values (i.e., a statement is either true or false in comparison to a set of assumptions), in deontic logics our statements characterize a fact as obligatory, prohibited or permitted. We therefore evaluate deontic statements in terms of validity and not of truth, and we evaluate facts in terms of legitimacy in relation to a deontic statement. For example, if we have ${\displaystyle [O]A}$ and we have ${\displaystyle \neg A}$, we do not assess ${\displaystyle \neg A}$ as ``false'' but rather as ``breach''.

\end{Definition*}

\sectopic{(5) Rule languages: }
Early work in RE focused on representing legal requirements as production rule models~\cite{maxwell2009developing}. Among the general-purpose rule languages developed for legal knowledge we mention \textbf{LegalRuleML}, a comprehensive XML-based representation framework for modeling and exchanging normative rules
~\cite{Palmirani18}. 
The following excerpt describes 
the compliance rule in LegalRuleML:

\begin{lstlisting}[language=XML, label={lst:lrml}, caption=Example of specifying a  compliance rule in LegalRuleML, mathescape=true]
<! -- Pre-condition requiring that the breach results in a non-unlikely risk to natural persons. -- >
<ruleml:And>
    <ruleml:Atom>
        <ruleml:Rel iri=":breach"/>
        <ruleml:Var>B</ruleml:Var>
    </ruleml:Atom>
    <ruleml:Atom>
        <ruleml:Rel iri=":riskToNaturalPersons"/>
        <ruleml:Var>R</ruleml:Var>
    </ruleml:Atom>
    <ruleml:Atom>
        <ruleml:Not>
        <ruleml:Rel iri=":unlikely"/>
        <ruleml:Var>B</ruleml:Var>
        <ruleml:Var>R</ruleml:Var>
        </ruleml:Not>
    </ruleml:Atom>
</ruleml:And>

<! -- constraint requiring that the notification N concerning the breach is either performed within 72 hours or is accompanied by the reasons for the delay. -- >
<ruleml:Atom>
    <not>
    <ruleml:Rel iri=":concerns"/>
    <ruleml:Var>N</ruleml:Var>
    <ruleml:Var>B</ruleml:Var>
    </not>
</ruleml:Atom>
<ruleml:Or>
    <ruleml:Atom>
        <ruleml:Rel iri=":delay"/>
        <ruleml:Var>N</ruleml:Var>
        <ruleml:lessthan>72</ruleml:lessthan>
    </ruleml:Atom>
    <ruleml:And>
        <ruleml:Atom>
            <ruleml:Rel iri=":ReasonsForDelay"/>
            <ruleml:Var>E</ruleml:Var>
        </ruleml:Atom>
        <ruleml:Atom>
            <ruleml:Rel iri=":contains"/>
            <ruleml:Var>N</ruleml:Var>
            <ruleml:Var>E</ruleml:Var>
        </ruleml:Atom>
    </ruleml:And>
</ruleml:Or>
\end{lstlisting}

\begin{figure}[H]
\centering

\includegraphics[width=0.8\linewidth] {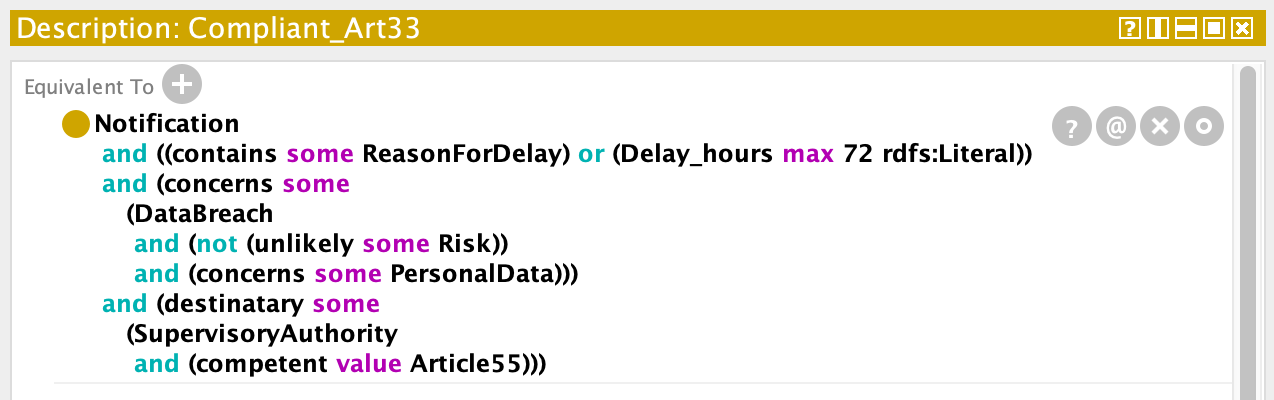}
\caption{An OWL axiom for \textit{Example}~\ref{eg1} as shown in the ontology editor Protégé.} \label{fig:protege_rule}
\vspace{-1em}
\end{figure}

\sectopic{(6) Semantic web-based methods: }
OWL and RDF 
are currently supported by semantic editors such as Protégé and semantic reasoners such as Pellet, FaCT++ and HermiT.
Some approaches in the existing literature~\cite{Francesconi22,AlKhalil16} propose specifying compliance criteria through an OWL ontology. 
Fig.~\ref{fig:protege_rule} shows a representation of the rule as derived from the ontology of domain concepts presented in Fig.~\ref{fig:ontology}. 

Following is the same information in OWL/XML Syntax, reusing the representation provided in Representation \ref{lst:owlr}:
\begin{lstlisting}[language=XML, label={lst2:lrml}, caption=Representation of \textit{Example}~\ref{eg1} in OWL/XML, mathescape=true]
<! -- constraint requiring that the notification concerning the breach is either performed within 72 hours or is accompanied by the reasons for the delay. -- >

    <EquivalentClasses>
        <Class IRI="#Compliant_Art33"/>
        <ObjectIntersectionOf>
            <Class IRI="#Notification"/>
            <ObjectUnionOf>
                <ObjectSomeValuesFrom>
                    <ObjectProperty IRI="#contains"/>
                    <Class IRI="#ReasonForDelay"/>
                </ObjectSomeValuesFrom>
                <DataMaxCardinality cardinality="72">
                    <DataProperty IRI="#Delay_hours"/>
                </DataMaxCardinality>
            </ObjectUnionOf>

<! -- constraint requiring that destinatary of the notification is the competent authority under Article 55. -- >
            <ObjectSomeValuesFrom>
                <ObjectProperty IRI="#destinatary"/>
                <ObjectIntersectionOf>
                    <Class IRI="#SupervisoryAuthority"/>
                    <ObjectHasValue>
                        <ObjectProperty IRI="#competent"/>
                        <NamedIndividual IRI="#Article55"/>
                    </ObjectHasValue>
                </ObjectIntersectionOf>
            </ObjectSomeValuesFrom>
\end{lstlisting}

The latest W3C standard for the semantic web, \textbf{SHACL} (Shapes Constraint Language), was created for validating the contents of an RDF-style graph database and has been used with promising results for representing compliance criteria. 
It allows to perform \textbf{SPARQL} queries (SPARQL will be introduced in section 3) assigning priorities to them (execution orders), a functionality that was missing from previous work on SPARQL~\cite{robaldo2022taking}.

%% file: compliance.tex
\section{From Representation to Compliance}
\label{sec:compliance}
\renewcommand{\lstlistingname}{Code \faIcon{code}}


Below, we focus on automated approaches that are proposed to assist requirements engineers in better understanding the regulation and help them capture a complete set of legal requirements. While not all of the representations discussed in Section~\ref{sec:representation} are referred to in this section, representing legal requirements is a key step towards enabling the automation strategies described below. 
\subsection{Improved Understanding of Regulation}~\label{subsec:understanding}
During requirements elicitation, requirements engineers often need to extract compliance-relevant information from different applicable regulations. Navigating through regulations manually is time-consuming and error-prone. Automated support for retrieving relevant information is thus advantageous.  
In RE, different approaches exist to enable querying regulations, e.g.,
~\cite{Sleimi:19}. 

We describe below a recent approach, referred to as \texttt{COREQQA}, which  leverages large language models (LLMs) readily fine-tuned for question answering (QA) to retrieve compliance-relevant information from regulations~\cite{abualhaija2022coreqqa,9920030}. To obtain a more comprehensive understanding of compliance, COREQQA can be adapted to answer questions from sources other than regulations, e.g., technical guidelines.  

\begin{good-to-know*}{}
  Generally speaking, QA follows two major paradigms in the NLP literature: information retrieval (IR)-based QA which relies on IR technologies to retrieve for a given question (also called \textit{query}) the most relevant document(s) or text passage(s) from a collection of documents, whereas machine reading comprehension (MRC)-based QA 
  relies on semantic analysis methods to demarcate in a given text passage the right answer for a given question. An end-to-end QA system should subsequently perform the two tasks, namely retrieving the most relevant text segment and extracting the likely answer therein.    
\end{good-to-know*}

%
\texttt{COREQQA} takes as input a question ($\mathcal{Q}$) posed in natural language (NL), and a regulatory document ($\mathcal{D}$). Using cosine similarity between dense representations of $\mathcal{Q}$ and partitions of $\mathcal{D}$, \texttt{COREQQA} then finds 
the likely answer to $\mathcal{Q}$ in $\mathcal{D}$ and returns that answer as output.  \texttt{COREQQA} consists of the following steps:

\noindent \textit{(1) Preprocess Text:} \texttt{COREQQA} processes $\mathcal{Q}$ and $\mathcal{D}$ using a simple NLP pipeline that decomposes the text into different tokens and sentences and further removes very frequent words such as prepositions. 
In the same step, \texttt{COREQQA} further partitions $\mathcal{D}$ into \textit{context spans}. The rationale behind partitioning is two-fold. First, the NLP technologies applied for extracting the likely answer have the limitation of processing at most 512 tokens of text. Second, such partitions contribute to the final output which can be manually reviewed and should therefore be of a reasonable size. The 
resulting list of context spans is passed on to step~(2).

\begin{Definition*}{}
  A \textit{context span} represents consecutive text that  collectively contains less than 512 tokens corresponding to an article in $\mathcal{D}$ or a partition thereof.
\end{Definition*}

\noindent \textit{(2) Select Context Span:} In the second step, \texttt{COREQQA} applies IR-based techniques to retrieve the list of context spans that are most relevant to the input question ($\mathcal{Q}$). The landscape of IR largely depends on semantic search methods, currently dominated by LLMs, which 
enable semantic-aware retrieval beyond mere textual overlap.  
\texttt{COREQQA} applies BERT cross-encoder-based similarity (BCE) which measures the semantic similarity of two text segments. BCE returns a score between zero and one, where zero indicates dissimilar and one indicates identical. Since a context span typically consists of multiple sentences, \texttt{COREQQA} computes the BCE score between $\mathcal{Q}$ and each sentence in the span.   

\begin{tip*}{}
Okapi Best Matching (BM25) was the best IR method for a long time~\cite{Whissell:11}. 
BM25, which is an improvement of the well-known term frequency - inverse document frequency (TF-IDF) method~\cite{McGill:83}, optimizes the weights of the terms using  probabilistic model and relevance feedback. Recent work in IR proposes a re-ranking strategy that combines BM25 with LLMs. You can try this out!
\end{tip*}

To illustrate, consider the example code~\ref{lst:bce}. Let $\mathcal{Q}$ be ``How should we handle personal data breach?'' (\texttt{Line 6}) and $\mathcal{D}$ be GDPR. For the sake of example, we show one context span (referred to as C) corresponding to Article 33(1) in GDPR, see \textit{Example}~1 in Section~\ref{sec:introduction}. The span consists of two sentences shown on \texttt{Line 8}. 
To measure the relevance of this context span to $\mathcal{Q}$, we compute BCE between $\mathcal{Q}$ and each sentence in C. We first define which LLM will be used to compute similarity on \texttt{Line 11}. We then feed in the inputs to the selected model (\texttt{Line 13}) and let the model predict the semantic similarity (\texttt{Line 14}). Finally, we sort the scores in descending order (\texttt{Line 17}) and take the maximum score as the relevance between $\mathcal{Q}$ and C on \texttt{Line 19}. To better understand the intuition behind taking the maximum instead of the average score as the final relevance value, we provide a code snippet (\texttt{Lines 22 - 24}) that displays BCE scores between $\mathcal{Q}$ and each sentence in C. 
In the example code~\ref{lst:bce}, $\mathcal{Q}$ has a high score with the first sentence, yet a very low score with the second sentence. Averaging the scores would reduce the likelihood that this context span be considered relevant.  
Depending on the pre-defined value K, the 
list of top-K relevant context spans are passed on to step~(3). 

\begin{lstlisting}[language=Python, label={lst:bce}, caption=Computing BCE between $\mathcal{Q}$ and a context span $C$, mathescape=true]
# Import the library
!pip install -U sentence-transformers
from sentence_transformers import CrossEncoder

# Define an input question $\mathcal{Q}$
Q = 'How should we handle personal data breach?'
# Assuming that $\mathcal{R}$ is partitioned into a list of context spans C, and each span is further split into sentences
C= ['In the case of a personal data breach, the controller shall without undue delay and, where feasible, not later than 72 hours after having become aware of it, notify the personal data breach to the supervisory authority competent in accordance with Article 55, unless the personal data breach is unlikely to result in a risk to the rights and freedoms of natural persons.',
'Where the notification to the supervisory authority is not made within 72 hours, it shall be accompanied by reasons for the delay.']
# Define the LLM to use for computing the similarity
model = CrossEncoder('cross-encoder/ms-marco-TinyBERT-L-2')
# Define the inputs for the model, including $\mathcal{Q}$ and a sentence in C
model_inputs = [[Q, span] for span in C]
scores = model.predict(model_inputs)
results = [{'input': inp, 'score': score} for inp, score in zip(model_inputs, scores)]
# Sort the scores in descending order 
results = sorted(results, key=lambda x: x['score'], reverse=True)
# The relevance between $\mathcal{Q}$ and C is the maximum score
relevance = list(values()[0])[1]

# Display the results for each sentence
print('The input question "'+Q+'"')
for item in results:
   print(' has a BCE score of '+str(list(item.values())[1])+' with the sentence "'+list(item.values())[0][1]+'"')
\end{lstlisting}


\noindent \textit{(3) Extract Answer:} \texttt{COREQQA} extracts a likely answer to $\mathcal{Q}$ from each context span in the top-K relevant spans. To do so, \texttt{COREQQA} utilizes LLMs which are readily fine-tuned to solve the QA task (more concretely, the MRC task).  A fine-tuned LLM takes as input $\mathcal{Q}$ and a context span and returns as output the likely answer to $\mathcal{Q}$ in that span. To illustrate, consider the example code~\ref{lst:roberta-qa}. Once the necessary libraries are imported (\texttt{Line 3}), the fine-tuned QA model can be specified (\texttt{Line 5}) and applied through the QA pipeline available in the Transformers library (\texttt{Lines 7 - 15}). Finally, the predicted answer can be displayed (\texttt{Line 18}). 

\begin{lstlisting}[language=Python, label={lst:roberta-qa}, caption=Extracting from a context span the likely answer to $\mathcal{Q}$, mathescape=true]
# Import the necessary libraries
!pip install transformers 
from transformers import AutoModelForQuestionAnswering, AutoTokenizer, pipeline
# Specify the QA model
model_name = "deepset/roberta-base-squad2-distilled"
# Instantiate pipeline
nlp = pipeline('question-answering', model=model_name, tokenizer=model_name)
# Specify the question and a context span
Q = 'How should we handle personal data breach?'
span = 'In the case of a personal data breach, the controller shall without undue delay and, where feasible, not later than 72 hours after having become aware of it, notify the personal data breach to the supervisory authority competent in accordance with Article 55, unless the personal data breach is unlikely to result in a risk to the rights and freedoms of natural persons. Where the notification to the supervisory authority is not made within 72 hours, it shall be accompanied by reasons for the delay.'
# Generate predictions
preds = nlp(
    question = Q,
    context = span,
)
# Display the results
print(
    f"answer: {preds['answer']}"
)
\end{lstlisting}



\subsection{Automated Compliance Verification}

\begin{figure}[H]
\centering
\includegraphics[width=0.9\linewidth] {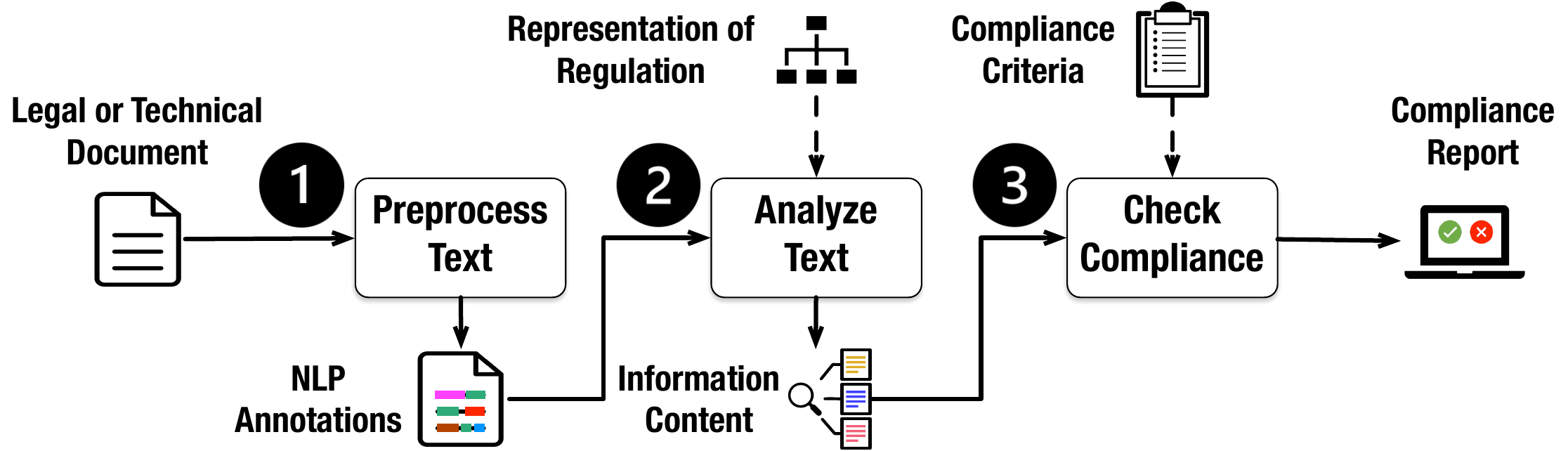}
\caption{General Pipeline for Automated Compliance Verification.} \label{fig:compliance}
\vspace{-1em}
\end{figure}

Fig.~\ref{fig:compliance} shows a general pipeline for automated compliance verification, which can be instantiated with different techniques. The approach takes a legal or technical document as input and returns a compliance report as output. The approach is composed of three main steps, steps \one -- \three in the figure elaborated below. 

\subsubsection{\one Preprocess Text. } 
This step applies an NLP pipeline to generate different annotations on the text in the input document. By annotations we mean information derived from the NLP tools about the text, such as part-of-speech tags, lemmas, named entities. 
We discuss below a comprehensive NLP pipeline for text preprocessing. For more details, we refer the reader to books such as~\cite{Jurafsky:09,Aggarwal:18}. 

\noindent \textit{(i) Tokenization} separates out tokens from running text. 

\noindent \textit{(ii) Sentence splitting} breaks up text into individual sentences. Detecting the beginning or ending of a sentence depends on some heuristics related to the language. For instance, a sentence in the English language starts with a capital letter and ends with a period. 

\noindent \textit{(iii) Part-of-Speech tagging} is the process of assigning a part-of-speech (POS) marker to each token in a given text sequence. Examples of POS tags include verbs like ``process'', nouns like ``data'', and adjectives like ``personal''. 

\noindent \textit{(iv) Lemmatization} involves identifying the canonical form (i.e., lemma) for each token in a given text sequence. For example, the lemma for both ``organization'' and ``organized'' is ``organize''. Lemmatization is used to normalize the text to enable more effective automated analysis such that the machine can learn common patterns despite the diversity across the documents of interest. 


\noindent \textit{(v) Constituency parsing} identifies the constituents that act as one structural unit, e.g., noun phrases and verb phrases. Constituency parsers generate a syntax (or parse) tree with the detailed POS tags of tokens which are further combined into phrasal structure, e.g., the noun phrase ``the personal data'' is composed of a definite article, an adjective, and a noun. 

\begin{good-to-know*}{}
  \textit{Text chunking}, similar to constituency parsing, divides the text into structural units albeit without performing detailed analysis on single words within a chunk. Thus, text chunking is computationally less expensive than constituency parsing.
\end{good-to-know*}

\noindent \textit{(vi) Dependency parsing} identifies the  binary grammatical relations in a given text, where the relation is directed from a dependent towards a governor. For instance, the verb ``notify'' in the phrase ``the controller shall notify'' has a directed relation ``nsubj'' with ``the controller'' indicating its subject.   

\noindent \textit{(vii) Named entity recognition} labels in a text the sequence of words which corresponds to a named entity such as the name of a person or organization.

\noindent \textit{(viii) Semantic role labeling} describes the semantic relationship between a verb and its related arguments, e.g., the verb ``to buy'' requires several arguments including at least ``the buyer'' and ``the item which is bought''.

\noindent \textit{(ix) Co-reference resolution} finds the mentions of the same entity in a document. 

The example code~\ref{lst:nlp-pipeline} applies the Stanza toolkit~\cite{stanza}, which is nowadays widely used to process NL text in Python. \textit{\textbf{Tip:} The example code uses the default pipeline. Try to rewrite it applying a customized pipeline!}
\begin{lstlisting}[language=Python, label={lst:nlp-pipeline}, caption=NLP Pipeline Demonstration, mathescape=true]
#Install and import the Stanza toolkit
!pip install stanza
import stanza
stanza.download('en') # download English model
nlp = stanza.Pipeline('en') # initialize English pipeline
doc = nlp("The processor shall assist the controller in ensuring the security of processing.") # run annotation over a sentence
print(doc)
print(doc.entities)
\end{lstlisting}


\begin{figure}[H]
\centering
\includegraphics[width=0.85\linewidth] {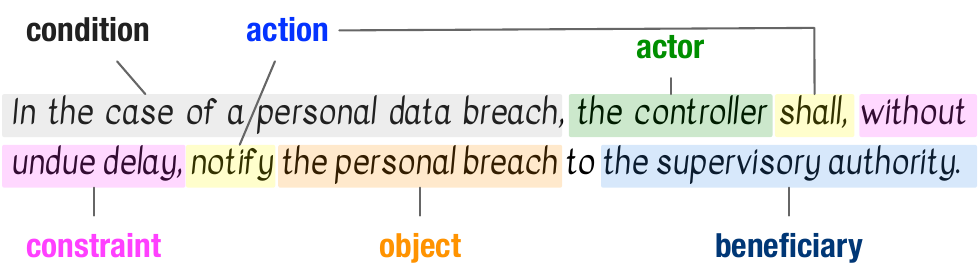}
\caption{Simplified illustration of legal SRs.} \label{fig:SR-REQ}
\vspace{-2em}
\end{figure}

\subsubsection{\two Analyze Text. }
This step performs in-depth syntactic and semantic analysis of the text in a given input document as a preparatory step towards checking its compliance against the applicable regulation. To this end, we discuss below two different approaches that have been proposed in the RE literature. 
In the remainder of this section, we use the following notation: Let $\mathcal{R}=\{r_1,\ldots,r_n\}$ and $\mathcal{T} = \{t_1, \ldots, t_m\}$ be the list of legal requirements derived from a regulation and the text segments in the input document, respectively. 

\textit{(A) Define/identify legal semantic roles: }
Legal semantic roles (also referred to as semantic metadata) identify text spans which hold certain meanings in the legal provision.  
Several of such roles have been proposed in the context of regulatory compliance in RE~\cite{amaral2023nlp,Bhatia:19,Sleimi18}, a subset of which is listed in Table~\ref{tab:SRs}.  

\begin{Definition*}{}
    \textit{A semantic role (SR)} describes the role of a participant with respect to a specific action. It is composed of a text span and a label. 
\end{Definition*}

Using the NLP annotations produced in the previous step, step~\two automatically identifies the legal semantic roles in each text segment $t_i$. The automated procedure utilizes primarily the syntactic analysis of a given text, such as POS tags and grammatical relations. It can also make use of the semantic roles identified by the NLP tools. Note that the NLP semantic roles cannot be one-to-one mapped to the legal semantic roles. The reason is that the former are more generic and describe any natural text, whereas the latter are specific to the legal domain. To identify SRs, existing approaches apply rules on the NLP annotations in combination with keywords (or markers) which enable distinguishing one SR from another. Table~\ref{tab:SRs} explains how each SR can be automatically identified. 
\begin{table}[t]
\caption{Legal SRs and their corresponding extraction rules~\cite{amaral2023nlp}.}\label{tab:SRs}
  \centering
	  \begin{threeparttable}[t]
\begin{tabularx}{0.98\textwidth} {@{} 
  p{0.15\textwidth} |X} 
\toprule
SR  & Extraction Rule\tnote{$\dag$} \\
\midrule
\textit{Action} & The \textit{action} is the \texttt{VP} that contains the root verb retrieved from the dependency parsing tree.  The action is the main verb in a statement. We note that a parsing tree contains only one root node, and this is inline with our goal of identifying one predicate for each statement.\\
\midrule
\textit{Actor} & The \textit{actor} is the \texttt{NP} containing the subject of the root verb in the statement (i.e., the \textit{action}). \\
\midrule
\textit{Object} & The \textit{object} is either the \texttt{NP} that contains the object of the root verb \textit{and} the \textit{action} does not contain any \textit{beneficiary} marker; \textit{or} the \texttt{VP} that starts with a preposition and is associated with the root verb.  \\
\midrule
\textit{Condition} & The \textit{condition} is the \texttt{PP}, \textit{or} \texttt{ADVP}, \textit{or} any phrase  annotated as subordinate clause by the dependency parser given that it contains a \textit{condition} marker such as if, once, in case of. \\
\midrule
\textit{Constraint} & The \textit{constraint} is identified in similar manner as \textit{condition} except that the phrase should contain a \textit{constraint} marker such as without, in accordance with, according to, unless. \\
\midrule
\textit{Beneficiary} & Similar to \textit{object}, the \textit{beneficiary}  is either the \texttt{NP} that contains a preposition linked to the root verb \textit{and} the \textit{action} contains a \textit{beneficiary} marker, \textit{or} the \texttt{NP} that contains a subject of a verb that is different from the root verb.  Since the \textit{beneficiary} is an entity that benefits from the \textit{action}, it can be associated directly with the action given that the \textit{action} has a marker that pinpoints the necessity of a \textit{beneficiary}. Alternatively, the \textit{beneficiary} can be an entity associated with another action in the statement. \\
\bottomrule
\end{tabularx}
\begin{tablenotes}
    \item[$\dag$] \texttt{VP}: verb phrase, \texttt{NP}: noun phrase, \texttt{PP}: prepositional phrase, \texttt{ADVP}: adverbial phrase.
\end{tablenotes}
    \end{threeparttable}
    \vspace{-2em}
\end{table}
\textit{The intermediary output} of step~\two in this case includes the legal semantic roles identified in each text segment of the input document, as depicted in Fig.~\ref{fig:SR-REQ}.

\textit{(B) Classify Text: }
One can analyze the text in the input document according to a predefined representation (e.g., conceptual model). In this case, after partitioning the text into segments considering the desired level of granularity (e.g., sentences), the resulting text segments can be classified according to the concepts in the representation. 
We explain next an existing approach in RE which classifies the textual content of privacy policies according to a comprehensive conceptual model created from GDPR~\cite{Aramal21,Torre20}. Several methods for creating representations, e.g., conceptual models, have been explained in Section~\ref{sec:representation}.
Let $\{t_1, \ldots, t_m\}$ be the list of text segments in the input privacy policy, and $\{c_1, \ldots, c_x\}$ be the list of concepts in the conceptual model. To illustrate, consider the following example sentence in a given privacy policy: \textit{``You can \textcolor{magenta}{update your information} in your profile or \textcolor{blue}{delete your data} by closing your account.''} This sentence can be classified as containing both the \textcolor{magenta}{\textsc{Right to rectify}} and \textcolor{blue}{\textsc{Right to remove}} personal data since the sentence discusses both concepts. This example shows the need for multi-label classification as it is often the case that any $t_i$ in the privacy policy may be relevant to multiple concepts at the same time. 
Specifically, the approach trains multiple binary classifiers, one for each concept in the representation. For the previous example, two binary classifiers will be created: one for predicting whether the sentence is about \textsc{Right to rectify} or \textit{not} \textsc{Right to rectify}, and another for predicting \textsc{Right to remove} or \textit{not} \textsc{Right to remove}.  


Selecting appropriate classification methods largely depends on the size of the training set. 
For dealing with the complexity of  hierarchical conceptual models, existing work proposes a hybrid classification method~\cite{amaral20,Torre20} which classifies the text using a combination of ML, semantic similarity and keyword search. 

For the concepts that have a sufficiently large number of training examples, ML is used. 
In the case of privacy policies, one needs to have a large collection of policies that are manually analyzed according to the conceptual model. For building a binary classifier for a specific concept (e.g.,  
\textsc{Right to remove}), all sentences in the document collection that are manually labeled with this concept are used as positive examples, and all sentences that are not labeled with this concept are negative examples.  
To build the feature matrix necessary for training the classifier, each training example is first preprocessed using the NLP pipeline presented in step~\one and then transformed into a numeric vector that encapsulates the textual content. These numeric vectors corresponding to the training examples contain the features from which the classifiers learn to distinguish between the different labels. 
Deriving numeric vectors (also called embeddings) mostly relies on LLMs~\cite{Reimers:19}. 

Semantic similarity can be also used to perform classification~\cite{das2021,Reimers:19,thongtan2019}. Specifically, the training examples associated with a specific concept (e.g., \textsc{Right to remove}) are collected into one group which is labeled with that concept. The training examples are again preprocessed and transformed into numeric vectors. Then, a representative vector for each group is created by averaging all vectors in that group. To predict whether a new example belongs to the group or not, the semantic similarity is computed between this representative vector and the vector of the new example. If the similarity score is greater that a predefined threshold, then the new example is predicted to be about the concept associated with that group; otherwise the new example is not about the concept. 

Cosine similarity is by far the most common metric applied for measuring the semantic similarity between two text segments. It takes as input  numeric vectors which represent two text segments and produces as output a score between 0 and 1, indicating the semantic similarity between the two segments; where 1 indicates that the two segments are identical. 
\begin{Definition*}{}
\textit{Cosine similarity} is the angle between the numeric vectors (embeddings) representing two text segments, computed as $cosine(v,w)=\frac{\vec{v} \cdot \vec{w}}{\lvert v\rvert \lvert w\rvert}$.
\end{Definition*}

Finally, when there are not enough training examples, one can use keyword search to identify the likelihood of a sentence to be about certain concepts. With the current advances in NLP, using keywords is probably an outdated technique. Alternatively, one can apply recent technologies based on few-shot learning. If the text of an unseen example contains at least one keyword that is associated with a concept, then the example is predicted to be about that concept.  
For instance, the keywords highlighted in \textcolor{magenta}{magenta} and \textcolor{blue}{blue} in the example above can be considered as  indicators highlighting the two concepts discussed in the sentence.
Keyword search method can produce many false positives, i.e., sentences that are wrongly labeled with a concept due to the presence of keywords. 
%
The intermediary output of this step includes the predicted labels (concepts) assigned to the text segments in the input document. 

\vspace*{-.5em}
\subsubsection{\three Check Compliance. }
This step identifies in the input document potential \textit{breaches} that violate 
the applicable laws. While  step~\two is concerned with analyzing the text according to the representation created from the regulation, step \three applies the compliance criteria which are defined either directly from the regulation or based on the representation. See Section~\ref{sec:representation} for more details. 

\begin{Definition*}{}
Recall from Section~\ref{sec:introduction} that a \textit{breach} occurs when 
some of the legal requirements stipulated in the regulation are not fulfilled (or are violated). 
In this chapter, we focus on content-related requirements, i.e., the requirements about what a document (e.g., privacy policy) should contain in order to be compliant with the regulation (e.g., GDPR). 
\end{Definition*}

Various approaches have been proposed in RE to automatically check the compliance of a given document against the applicable regulation. Such approaches are intended to provide assistance to human analysts in the process of checking compliance. 
We discuss below different possible approaches~\cite{Aramal21,amaral2023nlp,ilyas2023}.

\textit{(1) Using semantic similarity at sentence level: }
The regulation can be represented as single textual requirements. A straightforward method is then to measure the semantic similarity between each legal requirement and each text segment in the input document. 
%
Algorithm~\ref{alg:relevance} shows the general procedure. 
A text segment $t_i$ is deemed relevant to a requirement $r_j$ if the semantic similarity between their equivalent numeric vectors is above a predefined threshold ($\theta$). This threshold can be optimized according to the application context. 
A prerequisite for running the algorithm is to preprocess both the input document and the legal requirements using NLP pipeline from step~\one. This algorithm produces as output the list of relevant pairs $\langle r,t\rangle$ of requirements and text segments. Violations can then be detected considering which requirements have no corresponding relevant text in the input document.

\renewcommand{\algorithmiccomment}[1]{\bgroup\hfill//~#1\egroup}
\renewcommand{\algorithmicensure}{\textbf{Output:}}

\setlength{\textfloatsep}{3pt}
\begin{algorithm}[t]
\caption{Relevance Detection}\label{alg:relevance}
\begin{algorithmic}[1]
\REQUIRE $\mathcal{T}$:~an input document; and $\mathcal{R}$:~a list of legal requirements
\ENSURE $Rel$:~a list of relevant pairs $\langle r,t\rangle$

\STATE $Rel \leftarrow \emptyset$ \COMMENT{\textcolor{blue}{Start with an empty list of relevant pairs}} 
\STATE Partition $\mathcal{T}$ into text segments $\{t_1, \ldots, t_m\}$

\FOR{$t_i \in \mathcal{T}$}
    \STATE $t_i \leftarrow$ \texttt{vectorize($t$)}\COMMENT{\textcolor{blue}{Get the numeric representation of $t_i$}}
    \FOR{$r_j \in \mathcal{R}$}
        \STATE $\vec{r_j} \leftarrow$ \texttt{vectorize($r_j$)}
        \STATE sim $\leftarrow$ \texttt{cosine($t_i$, $r_j$)}
        \IF[\textcolor{blue}{If the similarity score greater than a threshold}]{sim $\geq \theta$}
            \STATE Add $\langle r_j,t_i\rangle$ to $Rel$ 
        \ENDIF
    \ENDFOR
    \ENDFOR
\end{algorithmic}
\end{algorithm}

\textit{(2) Using semantic similarity at phrasal level: }
Considering what we discussed earlier about analyzing the text at the phrasal level using semantic roles (SRs) in step~\two, the legal requirements in the regulation can also be decomposed according to the same set of SRs used to analyze the text in the input document. 
Compliance checking can be carried out with the following steps:

\noindent (i) \textbf{Aligning SRs:} For each requirement $r_i$ and text segment $t_j$, identify common SRs in $r_i$ and $t_j$. 
For $t_j$ to be compliant with $r_i$, the former is expected to include the same (or more) roles as those found in $r_i$. 
However, only SRs in $r_i$ need to be checked against what has been identified in $t_j$. Any SR that is in $r_i$ and not in $t_j$ is considered \textit{MISSING} and can therefore lead to 
\rev{a breach}. Any SR that is in $t_j$ and not in $r_i$ is ignored since it is \textit{NOT REQUIRED} by the regulation. Any SR that is in both $r_i$ and $t_j$ will be then further checked in step (ii). 

\noindent (ii) \textbf{Matching text:} Compute a score using a similarity metric to match the text of the same SRs in $r_i$ and $t_j$. If the score exceeds a predefined threshold then the two texts are considered to match, otherwise not. For this step, it is beneficial to extend the existing text with synonyms and other related words to increase the likelihood of text matching. 

\noindent (iii) \textbf{Assigning scores:} Finally, a score is assigned to $t_j$ indicating to what extent it satisfies $r_i$. The score is computed  based on missing SRs or those whose text does not match. Again, if this score is above a predefined threshold, $t_j$ is predicted to be satisfying $r_i$. Overall, $r_i$ is violated if there is no $t_j$ that  is predicted to be satisfying $r_i$.

\newcommand{\cmark}{\ding{51}}%
\newcommand{\xmark}{\ding{55}}%
\newcommand{\done}{\rlap{$\square$}{\raisebox{2pt}{\large\hspace{1pt}\cmark}}%
\hspace{-2.5pt}}
\newcommand{\wontfix}{\rlap{$\square$}{\large\hspace{1pt}\xmark}}

\begin{table*}[t]
\caption{Example of compliance checking at phrasal level.}\label{tab:sf-eg}
  \centering
\begin{tabularx}{\textwidth} {@{} 
  p{0.03\textwidth} |X} 
  \toprule
   $r_i$. & In the case of a personal data breach, the controller shall, without undue delay, notify the personal data breach to the supervisory authority.  \\
   \cmidrule{2-2}
   & \textit{condition}: \colorbox{red!25}{In the case of a personal data breach} \\
    & \textit{actor}: the controller \\
    &  \textit{action}: shall notify \\
     &\textit{constraint}:\colorbox{red!25}{without undue delay} \\
     &\textit{beneficiary}: the supervisory authority \\
    &\textit{object}: the personal data breach \\
   \midrule
     $t_j$. & ORGANIZATION X will inform the supervisory authority about data breach to facilitate conducting the right procedure.   \\
\cmidrule{2-2}
& 
    \begin{itemize}
    \item[\wontfix]  \textit{condition}: \colorbox{red!25}{MISSING}
    \item[\done] \textit{actor}: ORGANIZATION X
    \item[\done] \textit{action}: will inform
    \item[\wontfix]  \textit{constraint}: \colorbox{red!25}{MISSING}
    \item[\done] \textit{beneficiary}: the supervisory authority
    \item[\done] \textit{object}: data breach 
    \item[] \textit{reason}: to facilitate conducting the right procedure \colorbox{gray!25}{NOT REQUIRED}
    \end{itemize}
\\   
    \bottomrule     
    \end{tabularx}
\end{table*}
Table~\ref{tab:sf-eg} shows a simplified example where the compliance checking is performed at phrasal level between the requirement $r_i$ and a text segment $t_j$. By aligning the SRs, we immediately see that the SRs \textit{condition} and \textit{constraint} are missing since they are required in $r_i$, and yet not identified in $t_j$. Recall that our objective is to check the compliance of content-related requirements and thus, in this case, a missing condition is a breach. 
We also see that \textit{reason} is an additional SR in $t_j$ which does not have value for compliance since it is not required (i.e., not in $r_i$). 
Regarding matching text (step (ii) above), we note that matching the text spans of the SR \textit{actor} would require human input about \textit{ORGANIZATION X} being a controller. Matching the spans of \textit{action} would require extending the text in $t_j$ (``shall inform'') with synonyms (e.g., "notify" is a synonym of "inform"). Matching the spans for \textit{beneficiary} and \textit{object} is more straightforward. 
Four out of six SRs in $r_i$ are found in $t_j$, namely \textit{actor}, \textit{action}, \textit{beneficiary}, and \textit{object}. Thus, the score assigned to $t_j$ is $\frac{4}{6}=0.67$.

\newlist{todolist}{itemize}{2}
\setlist[todolist]{label=$\square$}

\textit{(2) Using rules combined with ML predictions: }
For the approach that utilizes ML for predicting concepts (according to some conceptual model), compliance can be checked by making sure all required concepts are predicted to be in the text according to the predefined set of compliance criteria. 
To illustrate, consider \textit{Example}~1 in Section~\ref{sec:introduction}. If the ML predictions contain \textsc{data breach notification}, but do not contain the \textsc{time limit} (``within 72 hours''), then 
\rev{a breach} is detected since the content of the input document is incomplete according to the GDPR provisions.

\subsection{Evaluation of Compliance Verification Solutions}
\label{subsec:evaluation}

For evaluating a compliance checking approach, one should assess how well the approach detects 
genuine breaches
in a given document, in which case true positives (TPs) represent 
the breaches that are correctly detected by the approach, false positives (FPs) represent 
the breaches that are incorrectly detected by the approach, and false negatives (FNs) represent 
the breaches that are missed by the approach.  Following this, evaluation metrics like precision (P) and recall (R) can be reported, computed as $P=\frac{TP}{TP+FP}$ and $R=\frac{TP}{TP+FN}$.

\begin{good-to-know*}{}
  FNs occur when the automated approach incorrectly identifies textual content as satisfying a legal requirement, i.e., missing 
  a breach.
  FPs, on the other hand, occur when the approach misses textual content that satisfies the legal requirement, thus incorrectly detecting a violation. 
\end{good-to-know*}

In practice, favoring precision or recall 
depends  on the manual effort which the human analyst will invest to correct these errors and   the (legal) consequences of missing 
a breach. For instance, reviewing incorrectly identified relevant text (FNs) takes less time and effort than reading through the entire document to find missed relevant text (FPs). Unlike the typical tendency in RE to favor recall, precision takes precedence in the context of regulatory compliance.

%% file: challenges.tex
\section{Current Challenges of Legal Requirements Analysis} \label{challenges}
Regulatory compliance and legal requirements analysis are long-standing research problems. Existing work has, however, barely scratched the surface when dealing with the inherent complexity of the legal domain, posing in turn challenges as we describe below.  

\sectopic{Defining the source landscape. }
The solutions discussed in this chapter considered only one regulation as the source from which legal requirements are elicited and analyzed. This, however, is not always sufficient. For instance, answering a question about data breach (see Section~\ref{subsec:understanding}) would typically require searching in multiple relevant regulations including, but not limited to, GDPR. Moreover, the abundance of cross-references pointing to external regulations indicates the strong inter-relation that exists in every legal system between regulations. We note that the solutions in this chapter can still be applicable in the case of multiple regulations only if these regulations are considered separately. Considering multiple regulations (or multiple versions of the same regulation) for building one comprehensive solution is still an open issue for future research.
    

\sectopic{Dealing with the temporal parameters of the law. }
Whether for understanding the law or checking compliance, accounting for the time when a regulation is in effect is paramount for identifying the applicable norms. 
Existing work in RE focuses on analyzing provisions at a certain time, which results in nothing more than a mere snapshot of the applicable norms. However, the regulatory framework changes over time, and capturing this change is a prerequisite for ensuring the continuity of the system's compliance.  
Multiple temporal dimensions exist in law, with the most important distinction being between the lifecycle of the legislative document, the time when the law is in force, and the time when it is applicable \cite{palmirani10}. 
Unless these temporal dimensions are taken into account in the representation, any resulting solution is not guaranteed to apply correctly, especially on and around the days when new regulation comes into force.
While legal experts can provide information about these temporal dimensions, (semi-)automatically capturing the regulatory changes over time is an under-investigated topic in RE. 

\sectopic{The trade-off between representations. }
As we have seen in this chapter, law and software engineering have different scopes and perspectives. There is a knowledge gap between norms and requirements, and current attempts at machine-analyzable representations of regulations are aimed at circumventing (with an ad-hoc approach) or filling (with a theoretical approach) that gap.
    When relying on ad-hoc approaches common in RE, selecting which representation method and which automated solution to apply largely depends on the application context. There are very few comparative studies that provide guidelines and indicators on how these methods fare against one another on a similar problem. 
On the other hand, relying on a legal-theoretical approach by deriving a set of logical formulas from a legal provision (as described in Section \ref{sec:representation}), guarantees a standard representation and common understanding, and thus reusability and automation, 
demonstrating the potential of such an approach to bridge this knowledge gap between the two fields~\cite{palmirani09}. 
However, despite advances in AI, formal representations of the law must still be developed manually since it is a highly complex activity~\footnote{Expressing the full semantics of legal requirements requires advanced nonmonotonic logical formalization. Achieving a standard for such a formalization involves the creation of rule languages (see Section~\ref{sec:representation}). An even more complete solution for representing legal requirements involves argumentation theory~\cite{jureta21},  
which provides details about how could the reasoning have gone in a different direction and why~\cite{Ceci_Gordon_2012}.
}.
In the context of RE, the analysis of legal requirements usually relies on a 
rather simple representation, where the norms are represented as NL sentences or templates. All the complexities mentioned above are thus alleviated, and the requirements can be understood without the need to master either the law or formal reasoning. Compliance criteria can also be represented as legal requirements specific to the target application (the artifact being engineered), with their traceability to the source provisions, similar to the approaches described in Section~\ref{sec:compliance}. 


\sectopic{Explainability. }The solutions described in this chapter can provide explainability for their results, but this explaination is limited to the compliance of a software to the legal requirements, without taking into account 
the reasoning followed when creating such legal requirements as a representation of the applicable norms.
In order to achieve the latter, the process of requirements elicitation will need to be extended, to encompass also the operations performed by the legal experts when interpreting the regulation. It is a high-risk, high-reward endeavor: it would imply the introduction of a degree of formalization in the interpretation performed by legal experts, but it would enable a compliance checking output that goes beyond mere Boolean decisions (compliance/breach).

%% file: conclusion.tex
\section{Conclusion}


This chapter presented an end-to-end pipeline for analyzing legal requirements.  The main focus in the requirements engineering (RE) field is centered around supporting the compliance checking activities, which in turn help engineers develop  software systems that are compliant with the applicable laws. 

Starting from the regulation, 
the first step is to abstract and capture the legal knowledge in some machine-analyzable representation. We explained in this chapter several options for representing such legal knowledge, one of the most common ones of which is conceptual modeling. To create conceptual models, engineers need to define the entities and the relations between them in a close collaboration with legal experts. Using these concepts, one can define compliance criteria about what is required for the regulation to be satisfied. 

Once there is an adequate representation of legal knowledge, automated solutions can be devised to assist engineers (or human analysts in general) in performing several activities concerning regulatory compliance. We have discussed in this chapter two scenarios: (i) Retrieving compliance-relevant information via  question-answering techniques; and (ii) developing methods that provide automated assistance in checking compliance. Such methods typically leverage natural language processing (NLP) and machine learning (ML) to categorize the content of a legal document according to predefined concepts. 


Legal requirements analysis is an essential activity in RE which, despite advanced solutions proposed in the literature, still provides many open directions for future research.